\documentclass{iau} 
\usepackage{graphicx}

\title[Ripples in the Milky Way Disk] 
{The Vertical Displacement \\ of the Milky Way Disk}

\author[Heidi Jo Newberg \& Yan Xu]   
{Heidi Jo Newberg$^1$,
 \and Yan Xu$^{2,1}$}

\affiliation{$^1$Dept. of Physics, Applied Physics, and Astronomy, Rensselaer Polytechnic Institute, \\ 110 8th Street,
Troy, NY 12180, USA \\ email: {\tt newbeh@rpi.edu} \\ [\affilskip]
$^2$Key Laboratory of Optical Astronomy, National Astronomical Observatories, Chinese Academy of Sciences, \\ Datun Road 20A, Beijing 100012, PR China \\ email: {\tt xuyan@bao.ac.cn}}

\pubyear{2015}
\volume{xxx}  
\jname{Title of your IAU Symposium}
\editors{A.C. Editor, B.D. Editor \& C.E. Editor, eds.}
\begin{document}

\maketitle

\begin{abstract}
An oscillating vertical displacement of the Milky Way, with a wavelength of about 8 kpc and and amplitude of about
100 pc (increasing with distance from the Galactic center) is observed towards the Galactic anticenter.  These oscillations are thought to be the result of disk perturbations from dwarf satellites of the Milky Way.  They explain the Monoceros
Ring and could be related to Milky Way spiral structure.
\end{abstract}

\firstsection 
\section{Introduction}

It has been fourteen years since \cite[Newberg et al. (2002)]{nyetal02} used a $2.5^\circ$-wide stripe of Sloan Digital Sky 
Survey \cite[(SDSS; York et al. 2000)]{yetal00} data to discover that the Milky Way's stellar spheroid
was not a smooth power law distribution of stars, but was instead dominated by density substructures
that turned out to be tidal streams from dwarf galaxies that are in the process of being torn
apart as they fall into the Milky Way's gravitational potential \cite[(Bullock \& Johnston 2005)]{bj05}.  These tidal
streams are still being discovered in the outskirts of the Milky Way and also in other galaxies 
\cite[(Mart{\'{\i}}nez-Delgado et al.2010)]{MD10}.

Figure 1 of \cite[Newberg et al. (2002)]{nyetal02} shows the density of color-selected turnoff stars along the Celestial Equator
from the SDSS, showing structure that is now known as the Sgr dwarf tidal stream, the Virgo Overdensity, and the Monoceros
Ring.  The Monoceros Ring has been a particularly controversial structure originally identified at $(l,b)=(223^\circ,20^\circ)$ and 
mean apparent magnitude $g_0=19.4$, where the 
subscript ``0" indicates that the magnitude has been corrected for reddening.

If one looks in detail at the substructure near the anticenter in Figure 1 of \cite[Newberg et al. (2002)]{nyetal02}, there 
appears to be an oscillation in the
star counts above and below the Galactic plane that was unnoticed for more than a decade.  There are more stars above the
plane at $g_0=15$, more stars below the plane at $g_0=17.5$, more stars above the plane at $g_0=19$, and more stars
below the plane at $g_0=20$.  It turns out that this oscillation is real, is not due to the larger extinction in the south
Galactic cap, and is present over more than $100^\circ$ of Galactic longitude near the anticenter at low latitude.


\section{An Oscillating Vertical Displacement of the Milky Way Disk}

In \cite[(Xu et al. 2015)]{xn15}, we showed that the Galactic disk is not symmetric around the $b=0$ plane. As one looks towards the anticenter at low Galactic latitude the sign of the asymmetry oscillates with distance from the Sun.
There are more stars in the north at distances of about 2 kpc from the Sun, more stars in the south at 4-6 kpc from the Sun, 
more stars in the north at 8-10 kpc from the Sun, and possibly more stars in the south 12-16 kpc from the Sun.  The
asymmetry is observed in the Galactic longitude range $110^\circ < l < 229^\circ$.  The Galactocentric distance to the observed structure
increases slightly from the second quadrant to the third quadrant, roughly following the opening
of the Milky Way spiral arms.  We fit an exponential disk model with an oscillating, axisymmetric vertical displacement to the
star counts within 8 kpc of the Sun looking towards the Galactic anticenter; the best fit raises the disk midplane 
by 70 pc, 2.5 kpc from the Sun; and
lowers the disk midplane by 170 pc at 6 kpc from the Sun.  The amplitude of the oscillation
increases farther from the Galactic center.

These results were obtained from analysis of seven SDSS stripes that 
pass through the Galactic plane at constant Galactic longitude.  Photometric data with $10^\circ<|b|<30^\circ$ was
divided up into $2.5^\circ \times 2.5^\circ$ bins, positioned symmetrically on each side of the Galactic plane at $b=0^\circ$.
We created H-R diagrams for each sky patch.  We then subtracted the H-R diagram from that of the symmetric sky patch on the 
other side of the Galactic plane.  The result was an oscillating pattern of black and white main sequences.  At brighter
magnitudes (closer distances) there was a white main sequence, meaning there were more main sequence stars in the north.
Going fainter there was a dark main sequence (more stars in the south), then another white main sequence (more in the north), and
at the faintest portion of the subtractions there was possibly another darker main sequence.  Note that an error in reddening
correction could not cause this pattern because the reddening direction is roughly aligned with the main sequence, so
reddening would just move the stars along the main sequence and would not cause the main sequences to be misaligned
in magnitude from one hemisphere to the other.

Spectra of the stars in the brighter two main sequences are as expected for disk stars in kinematics, metallicity, and
density distribution.  They show the familiar asymmetric drift in the velocity distribution that is fit exactly, with no free
parameters, by the formulae in \cite[Sch{\"o}nrich (2012)]{s12}.  We did not have spectra for the stars in the outer two
structures at the positions of the Monoceros Ring and Triangulum-Andromeda Cloud, but presumably these are
also part of the Milky Way stellar disk and not separate tidal streams.  This is supported by recent claims that Tri-And
stars are disk-like \cite[(Price-Whelan et al.2015)]{pwetal15}.

The details of the analysis and results can be found in the \cite[Xu et al. (2015)]{xn15} paper.

\section{Implications}

Our result implies that the disk of the Milky Way extends out to 25 kpc from the Galactic center, and has oscillating
vertical displacements.  We suspect that previous studies showing the stellar disk density falling off abruptly 
at 15 kpc from the Galactic center were compromised by the unexpected shift in stars from one side of the Galactic plane to the other.
In our new picture of the disk, the Monoceros Ring and Triangulum-Andromeda Clouds are part of the stellar disk, and
{\it not} dwarf galaxy tidal streams.  These structures are
now presumed to be at peak displacements (one to the north and one to the south) of the disk midplane.

At present, there is only one mechanism known to cause vertical displacements in the disk midplane, and that is
the disk response to a dwarf galaxy (or dark subhalo) falling into the Milky Way.  A Sagittarius dwarf-sized galaxy
falling into the Milky Way produces qualitatively similar vertical displacements, though the results have not yet been
matched in detail \cite[(G{\'o}mez et al. 2013)]{getal13}. This same mechanism could also explain the observed 
oscillation of the stellar density perpendicular
to the plane \cite[(Yanny \& Gardner 2013)]{yg13}, and the coherent substructure in the velocities of disk stars that has been found in the SDSS 
\cite[(Widrow et al. 2012)]{wetal12}, RAVE \cite[(Williams et al. 2013)]{}, and
LAMOST \cite[(Carlin et al. 2013)]{} surveys.  Again, the observations have not been matched in detail with a particular model but qualitatively
they look similar.

Of particular interest is the relationship between wavelike oscillations observed in the Milky Way stellar disk and those 
observed in the gas.  Hydrodynamic simulations have been matched with the
observed gas oscillations to predict the approximate mass and orbit of a dwarf galaxy that is presumed to
have produced the disturbance \cite{cetal11}.  It is possible that the gas and stars are exhibiting wave 
motions from different satellites, since the physics of wave propogation and damping are different in these
two Galactic components.  However, this relationship remains to be fully tested.

Another open question is the relationship between the vertical displacement of the disk and spiral structure.  N-body 
simulations of satellites passing near the Milky Way disk
produce spiral structure.  However, spiral arms are very low scale height structures that
are populated by bright, young stars and star-forming regions that are thought to be the result of compression
of the gas in spiral density waves.  Spiral arms are not simply density variations in the typical disk star
population.  On the other hand, it is tempting to think that an infalling
dwarf galaxy could be the energy source that maintains spiral density structure over a large portion of the age of the 
Universe.  Satellites could perturb the disk on each orbital passage, explaining why most spiral galaxies are not
grand design spirals with easily traced spiral arms.

Vertical oscillations in the gas have been observed in galaxies outside of the Milky Way 
\cite[(Matthews \& Uson 2008)]{mu08}, but oscillations
in the stellar disks have not.  It is harder to observe oscillations in the stellar disk because the stars have a 
scale height that is larger than the observed amplitude of the oscillations.  They can only be seen in edge-on
galaxies, and the oscillations will be blurred along our line-of-sight through the galaxy.  However, we expect 
that if people look for vertical oscillations of the stars in external galaxies that they will find them.

For more than a decade, there has been a controversy over the identity of the Monoceros Ring.  Half of the
community (including me) thought that the Monoceros Ring was tidal debris from an infalling dwarf galaxy.  Half 
of the community thought that the Monoceros Ring was a warping or flaring of
the stars in the disk.  Our result suggests that the Monoceros Ring is neither; it is a wavelike perturbation of
the disk caused by a satellite (dwarf galaxy or dark subhalo) of the Milky Way.


\begin{thebibliography}{}


\bibitem[Bullock \& Johnston(2005)]{bj05} Bullock, J.~S., \& Johnston, K.~V.\ 2005, \textit{ApJ}, 635, 931 

\bibitem[Carlin et al.(2013)]{cetal13} Carlin, J.~L., DeLaunay, J., Newberg, H.~J., et al.\ 2013, \textit{ApJL}, 777, L5 

\bibitem[(Chakrabarti et al.2011)]{cetal11} Chakrabarti, S., Bigiel, F., Chang, P., \& Blitz, L.\ 2011, \textit{ApJ}, 743, 35 

\bibitem[G{\'o}mez et al.(2013)]{getal13} G{\'o}mez, F.~A., Minchev, I., O'Shea, B.~W., et al.\ 2013, \textit{MNRAS}, 429, 159 



\bibitem[Mart{\'{\i}}nez-Delgado et al.(2010)]{MDetal10} Mart{\'{\i}}nez-Delgado, D., Gabany, R.~J., Crawford, K., et al.\ 2010, \textit{AJ}, 140, 962 

\bibitem[Matthews \& Uson(2008)]{mu08} Matthews, L.~D., \& Uson, J.~M.\ 2008, \textit{ApJ}, 688, 237-244 

\bibitem[Newberg et al.(2002)]{nyetal02} Newberg, H.~J., Yanny, B., Rockosi, C., et al.\ 2002, \textit{ApJ}, 569, 245

\bibitem[Price-Whelan et al.(2015)]{pwetal15} Price-Whelan, A.~M., Johnston, K.~V., et al.\ 2015, \textit{MNRAS}, 452, 676 


\bibitem[Sch{\"o}nrich(2012)]{s12} Sch{\"o}nrich, R.\ 2012, \textit{MNRAS}, 427, 274 

\bibitem[Widrow et al.(2012)]{2012ApJ...750L..41W} Widrow, L.~M., Gardner, S., Yanny, B., Dodelson, S., \& Chen, H.-Y.\ 2012, \textit{ApJL}l, 750, L41

\bibitem[Williams et al.(2013)]{wetal13} Williams, M.~E.~K., Steinmetz, M., Binney, J., et al.\ 2013, \textit{MNRAS}, 436, 101 

\bibitem[Xu et al.(2015)]{2015ApJ...801..105X} Xu, Y., Newberg, H.~J., Carlin, J.~L., et al.\ 2015, \textit{ApJ}, 801, 105 

\bibitem[Yanny \& Gardner(2013)]{yg13} Yanny, B., \& Gardner, S.\ 2013, \textit{ApJ}, 777, 91 

\bibitem[York et al.(2000)]{2000AJ....120.1579Y} York, D.~G., Adelman, J., Anderson, J.~E., Jr., et al.\ 2000, \textit{AJ}, 120, 1579 

\end{thebibliography}
\end{document}